\documentclass[12pt,preprint]{emulateapj}
\usepackage{graphicx}

\shorttitle{Completeness of Imaging Surveys for Eccentric Exoplanets}
\shortauthors{Stephen R. Kane}
\slugcomment{Submitted for publication in the Astrophysical Journal}

\begin{document}

\title{Completeness of Imaging Surveys for Eccentric Exoplanets}
\author{Stephen R. Kane}
\affil{NASA Exoplanet Science Institute, Caltech, MS 100-22, 770
  South Wilson Avenue, Pasadena, CA 91125, USA}
\email{skane@ipac.caltech.edu}


\begin{abstract}

The detection of exoplanets through direct imaging has produced
numerous new positive identifications in recent years. The technique
is biased towards planets at wide separations due to the difficulty in
removing the stellar signature at small angular separations. Planets
in eccentric orbits will thus move in and out of the detectable region
around a star as a function of time. Here we use the known diversity
of orbital eccentricities to determine the range of orbits which may
lie beneath the detection threshold of current surveys. We quantify
the percentage of the orbit which yields a detectable signature as a
function of semi-major axis, eccentricity, and orbital inclination and
estimate the fraction of planets which likely remain hidden by the
flux of the host star.

\end{abstract}



\section{Introduction}
\label{intro}

The diversity of exoplanetary systems has been revealed through a
variety of detection techniques which are sensitive to different kinds
of planets. The majority of planets have been detected indirectly via
the radial velocity (RV) and transit methods, but the technique of
direct imaging has made important contributions by detecting young
planets at wide separations ($\gtrsim 10$~AU). Direct imaging is a
difficult technique to execute and requires the use of adaptive optics
and high contrast imaging to distinguish the flux of the planet from
that of the host star \citep{opp09}. The characteristic properties of
atmospherically-induced speckle noise can be treated using adaptive
optics techniques \citep{rac99,mac07}. Speckle noise due to
telescope/optics imperfections require more advanced techniques such
as angular differential imaging (ADI) \citep{mar05,mar06,laf07a},
spectral differential imaging (SDI) \citep{laf07b,vig10}, and
polarimetric differential imaging (PDI) \citep{qua12}. The list of
both technical and scientific achievements from exoplanet imaging
surveys is far too exhaustive to describe here in detail, but include
such milestone exoplanet discoveries as beta Pic b
\citep{lag10,cur11}, the HR~8799 planetary system
\citep{mar08,mar10,hin11,cur12a}, and Fomalhaut~b
\citep{kal08,cur12b}.

The star--planet separations explored by the imaging and RV techniques
are complementary to each other since the RV method is biased towards
shorter orbital periods. Although the imaging technique requires
considerable follow-up observations to constrain the orbits of the
planets detected, RV planets provide complete Keplerian orbital
solutions at the time of discovery which may be used to study the
overall statistics of exoplanetary systems. The eccentricity
distribution has been investigated for RV planets \citep{she08,hog10}
and also compared to that of the Kepler candidates
\citep{moo11,kan12}. The origin of the observed distribution has also
been studied in the context of planet formation scenarios and early
dynamical evolution \citep{nam05,rib07,for08,mal09,kle12}. If this
eccentricity distribution applies also to the kinds of planets
detected via direct imaging then the detectability of those planets
will depend on where they are in their orbit since the star--planet
separation is time dependent \citep{bra06,bon12}.

Here we investigate the effects of exoplanet orbital eccentricity on
the detection efficiency of imaging surveys. We describe the
constraints on observing planets in eccentric orbits depending upon
the sensitivity of direct imaging experiments to the inner region of
the system and the cadence of the observations. We calculate the
percentage of the orbit for which the planet will be detectable as a
function of eccentricity and the sensitivity threshold. We study the
known eccentricity distribution as a function of semi-major axis and
star--planet angular separation. Finally, we use this distribution to
estimate the detection efficiency of direct imaging for eccentric
exoplanets.


\section{Observational Constraints}
\label{observational}

In addition to the science requirements of imaging experiments for
achieving high-contrast observations that allow one to detect the
extreme star--planet flux ratio, one also needs to contend with the
limit of how close to the host star can reasonably be probed for
planets. The angular resolution needed for achieving this is limited
by the diffraction-limited resolution where the objects are separated
by $\lambda/D$ where $D$ is the aperture diameter of the telescope. In
practice, there are further techniques that may be brought to bear on
the problem that can further suppress the starlight and enhance the
contrast ratio. These techniques vary enormouslessly in both their
effectiveness and their number. Here we quantify the problem in terms
of a uniform inner region around a star for which planet detections
are inaccessible. We refer to this region as an {\em exclusion zone}
with radius $r_e$. Although a practical exclusion zone will be quite
complicated non-uniform pattern due to diffraction and speckle
effects, this allows us specify a lower-limit which may be adapted to
any particular experiment.

\begin{figure}
  \includegraphics[angle=270,width=8.2cm]{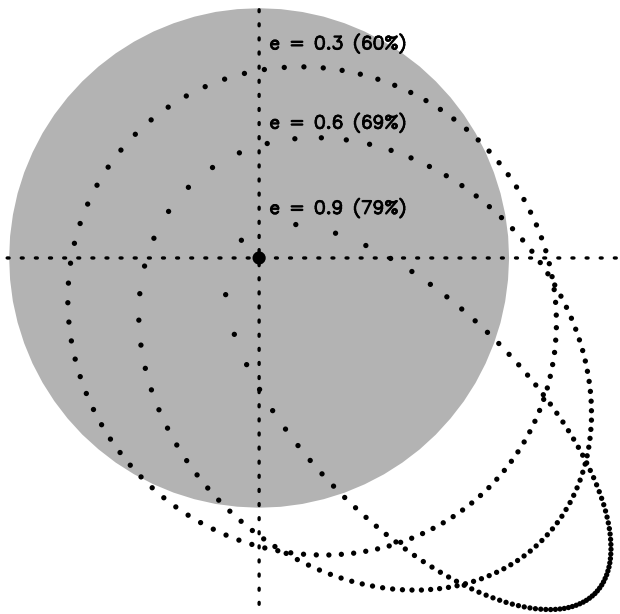}
  \caption{Three example orbits for a planet in orbit around a direct
    imaging target. The shaded region indicates the angular region
    around the star for which the example imaging experiment is
    insensitive to positive planet detections (exclusion zone). The
    orbits each have a semi-major axis equal to the radius of the
    exclusion zone and have eccentricites of 0.3, 0.6, and 0.9. The
    orbits are shown as points which indicate the planet location in
    equal time increments. The numbers in parentheses show the
    perecentage of the orbit that lies outside the exclusion zone.}
  \label{orbits}
\end{figure}

The problem of detecting eccentric exoplanets for which the
star--planet separation is time-dependent is portrayed in Figure
\ref{orbits}. The shaded region represents the exclusion zone for a
given star. The three orbits shown represent three Keplerian orbits
which have identical components with the exception of eccentricity,
shown here for $e = 0.3$, 0.6, and 0.9. The argument of periastron is
set to $\omega = 315\degr$ for this example. The semi-major axis in
all cases is set to $a = r_e$, such that a circular orbit would lie on
the threshold of detectability. The orbits are shown as points which
indicate the planet location in equal time increments to illustrate
the relative amount of time the planet spends at different phases of
its eccentric orbit. In this example, there are clear advantages for
highly eccentric planets whose orbits result in a larger angular
separation from the host star and a larger percentage of the orbit
outside of the exclusion zone.

\begin{figure}
  \includegraphics[angle=270,width=8.2cm]{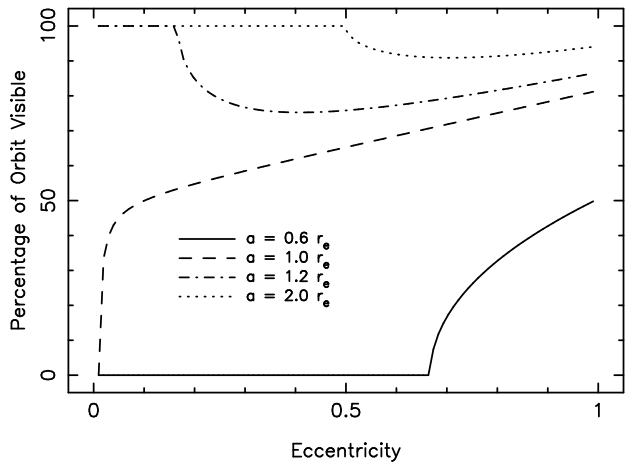}
  \caption{The dependence of the percentage of an orbit spent outside
    the exclusion zone on the orbital eccentricity for four different
    semi-major axis. The semi-major axes are in units of the exclusion
    zone radius, $r_e$.}
  \label{percentage}
\end{figure}

The percentage of the orbit during which the planet is in the
detectable region depends on the eccentricity and semi-major axis
relative to $r_e$. This is shown in Figure \ref{percentage} for four
different values of $a$ (in units of $r_e$). For $a < r_e$ there are
large swathes of eccentricities for which the orbit never exits the
exclusion zone. The limit at which the planet never exits the
exclusion zone is found from evaluating the star--planet separation at
apastron, that is where $a (1 + e) < r_e$. As $e$ approaches unity,
this limit is located at $a = 0.5 r_e$. A similar bound on a
semi-major axis beyond which the planet never enters the exclusion
zone is more open-ended since the periastron boundary condition of $a
(1 - e) > r_e$ approaches infinity as $e$ approaches unity.

Another observation constraint is that of the observing cadence. The
planet is moving at its slowest speed when the star--planet separation
is approaching maximum. Thus even sparse sampling should in principle
allow the tracking of the planetary orbit. There are however two
things to consider here. Firstly, if the movement of the planet is
less than the positional accuracy of the planetary
point-spread-function then this will create degeneracy in the derived
Keplerian solution for the planet. On the other hand, too sparse
sampling could result in the planet being missed in future (or past)
observations if the apastron is near the detection threshold, such as
the $a = 0.6 r_e$ case shown in Figure \ref{percentage}. Note that the
examples shown here aid in defining the problem posed by eccentric
orbits, but assume face-on orbits for the specific numbers
calculated. We discuss the effects of inclination in Section
\ref{deteff}.


\section{Eccentricity Distribution at Long Periods}
\label{eccdist}

Here we briefly discuss the eccentricity distribution of the known
exoplanets, particularly at relatively long periods. To do this, we
utilize the orbital parameters stored in the Exoplanet Data
Explorer\footnote{\tt http://exoplanets.org/} \citep{wri11}. We
extracted the data of 513 planets, along with the host star
properties, with data current as of 20th November 2012. These data
include all planets with the necessary Keplerian orbital solutions as
well as stellar distances.

\begin{figure*}
  \begin{center}
    \begin{tabular}{cc}
      \includegraphics[angle=270,width=8.2cm]{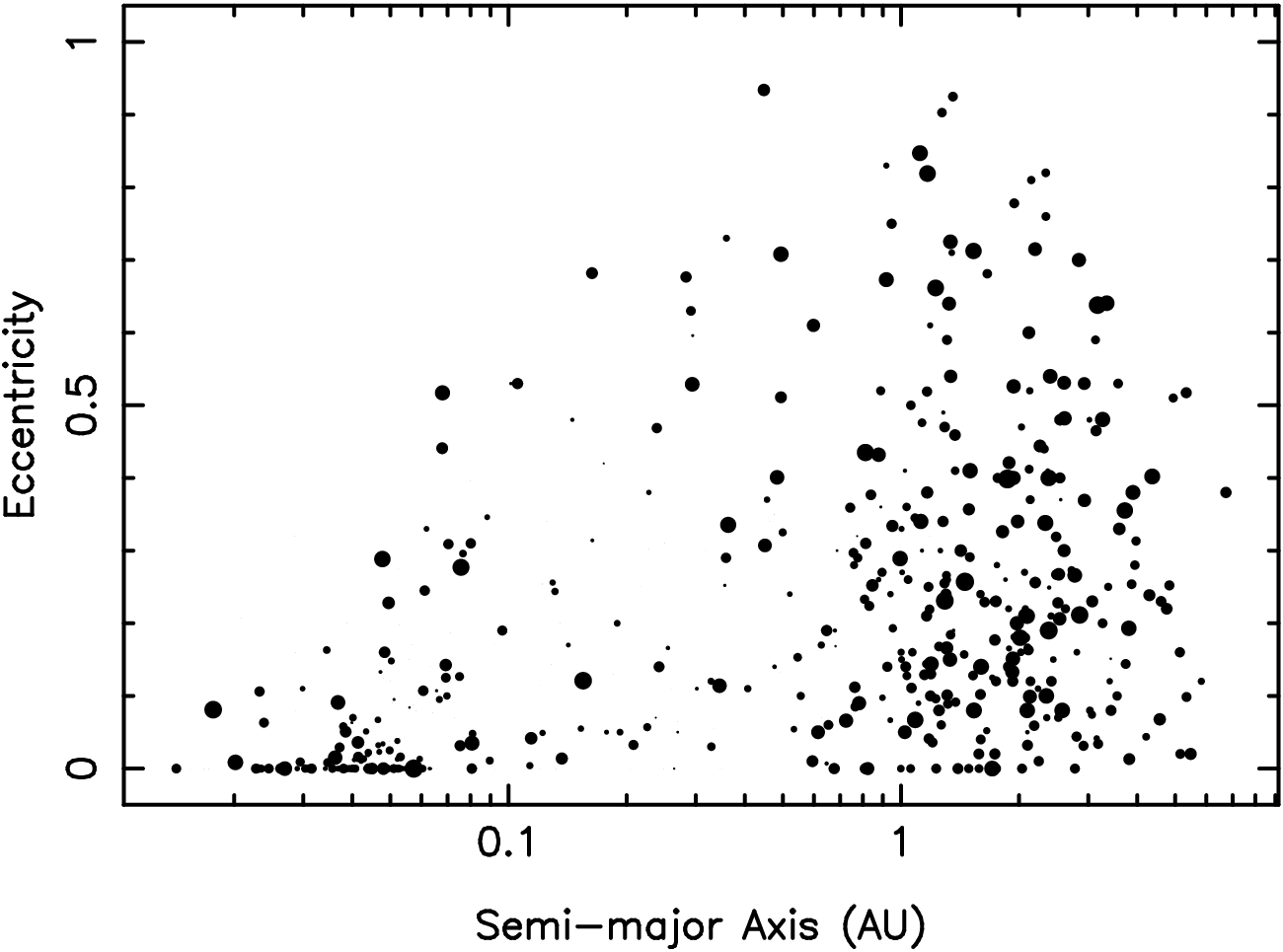} &
      \includegraphics[angle=270,width=8.2cm]{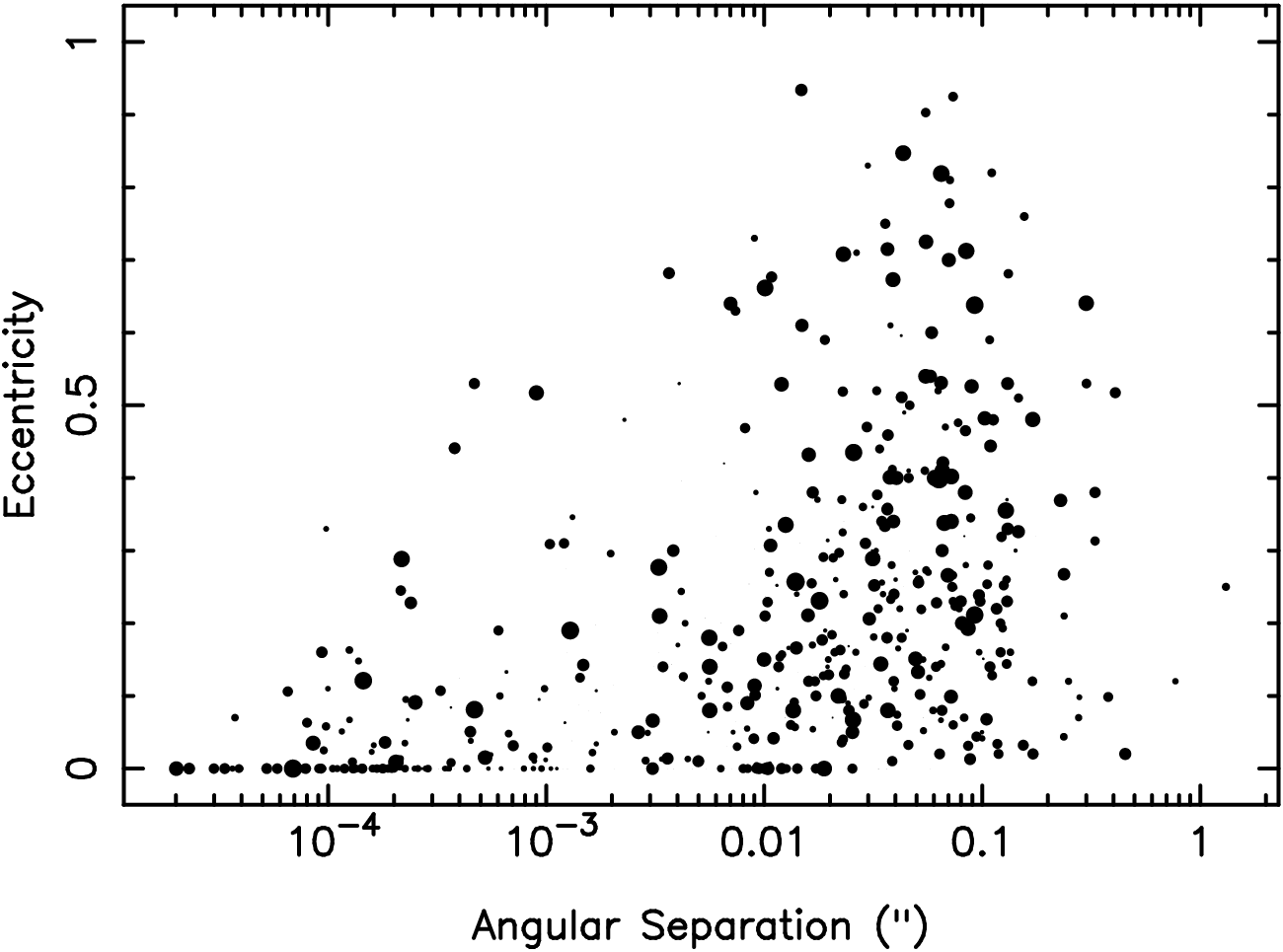}
    \end{tabular}
  \end{center}
  \caption{Left panel: The eccentricity of the known exoplanets as a
    function of semi-major axis. Right panel: The angular separation
    (arcsecs) of the same sample of planets at apastron and assuming a
    face-on orbit ($i = 0\degr$). In each plot, the size of each
    plotted point is logarithmically proportional to the mass of the
    planet.}
  \label{exoplanets}
\end{figure*}

In Figure \ref{exoplanets} we have plotted these data to show the
distributions of eccentricity values. The left plot shows this
distribution as a function of semi-major axis, where we have
logarithmically scaled the size of the points depending on the
planetary mass. The relative dearth of planets between 0.1~AU and
1.0~AU is well-known, noted for example by \citet{but06}. As is clear
from the plot, the mean eccentricity increases with semi-major axis
and is 0.06, 0.24, and 0.27 for semi-major axis ranges of 0.0--0.1,
0.1--1.0, and 1.0--2.0 respectively. What is not so clear is the
increase of planet mass with eccentricity; 2.5, 3.6, and 5.4 Jupiter
masses for eccentricity ranges of 0.0--0.3, 0.3--0.6, and 0.6--1.0
respectively. This is of particular relevance to the imaging surveys
since they are more sensitive to more massive planetary companions.

In the right panel of Figure \ref{exoplanets} we have plotted the
eccentricity against the calculated angular projected separation of
the planet from the star. This angular separation is calculated for a
face-on orbit when the planet is located at maximum separation
(apastron). For a Keplerian orbit, the time-dependent star--planet
separation, $r$, is given by
\begin{equation}
  r = \frac{a (1 - e^2)}{1 + e \cos f}
  \label{separation}
\end{equation}
where $f$ is the true anomaly. The general expression (in radians) for
the angular separation as a function of time is then
\begin{equation}
  \Delta \theta = \frac{r}{d} \left( \cos^2 (\omega + f) + \sin^2
  (\omega + f) \cos^2 i \right)^{\frac{1}{2}}
  \label{angsep}
\end{equation}
where $\omega$ is the periastron argument, $i$ is the orbital
inclination, and $d$ is the star--observer distance \citep{kan11}.
This angular separation is a useful diagnostic to show which of the
known exoplanets may yield positive results from a certain imaging
experiment. What is clear is that many of those planets which fall
into a detectable range of angular separations will be at the apastron
of an eccentric orbit.


\section{Detection Efficiency for Eccentric Orbits}
\label{deteff}

\begin{figure*}
  \begin{center}
    \begin{tabular}{ccc}
      \includegraphics[angle=270,width=5.2cm]{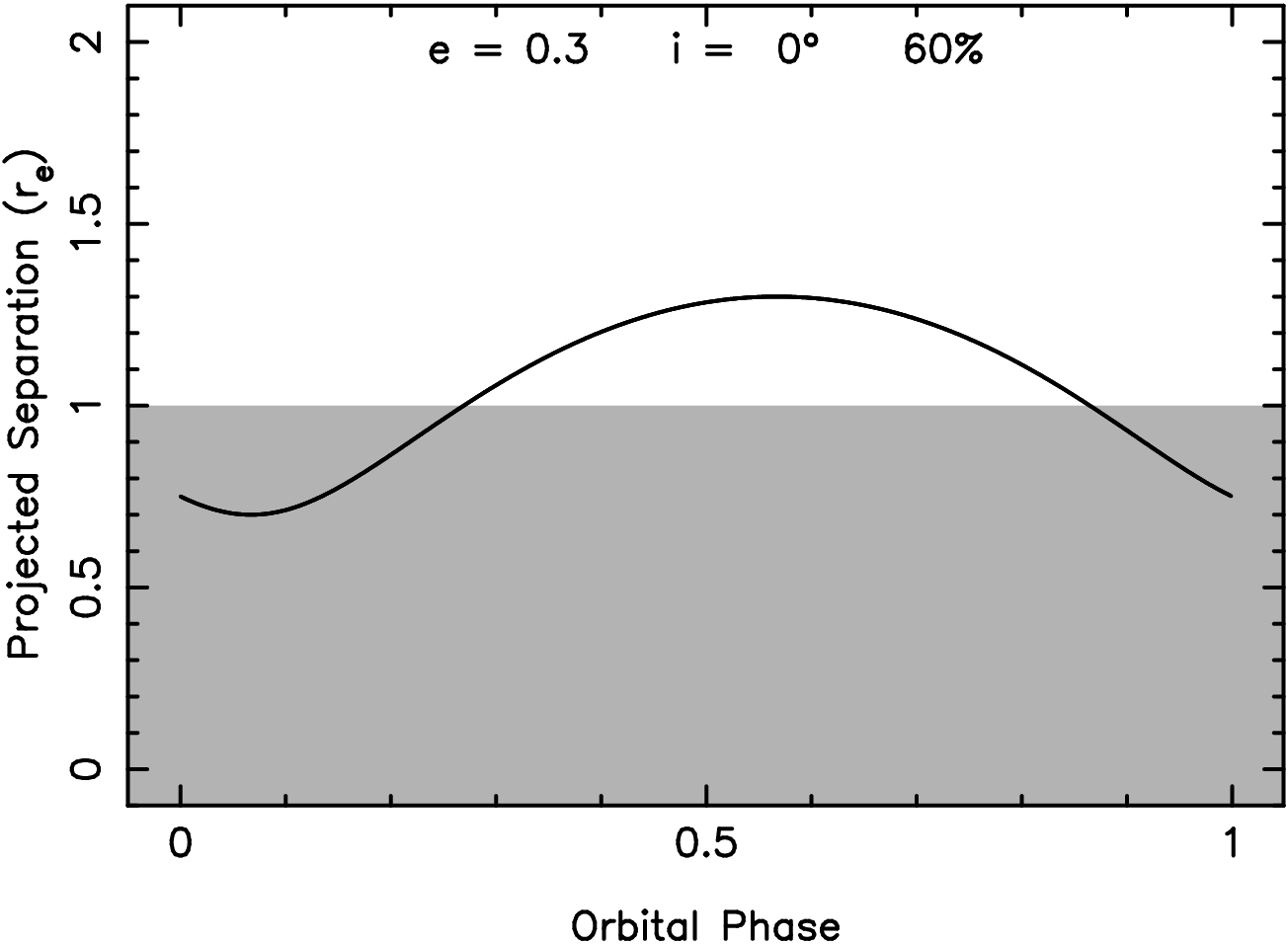} &
      \includegraphics[angle=270,width=5.2cm]{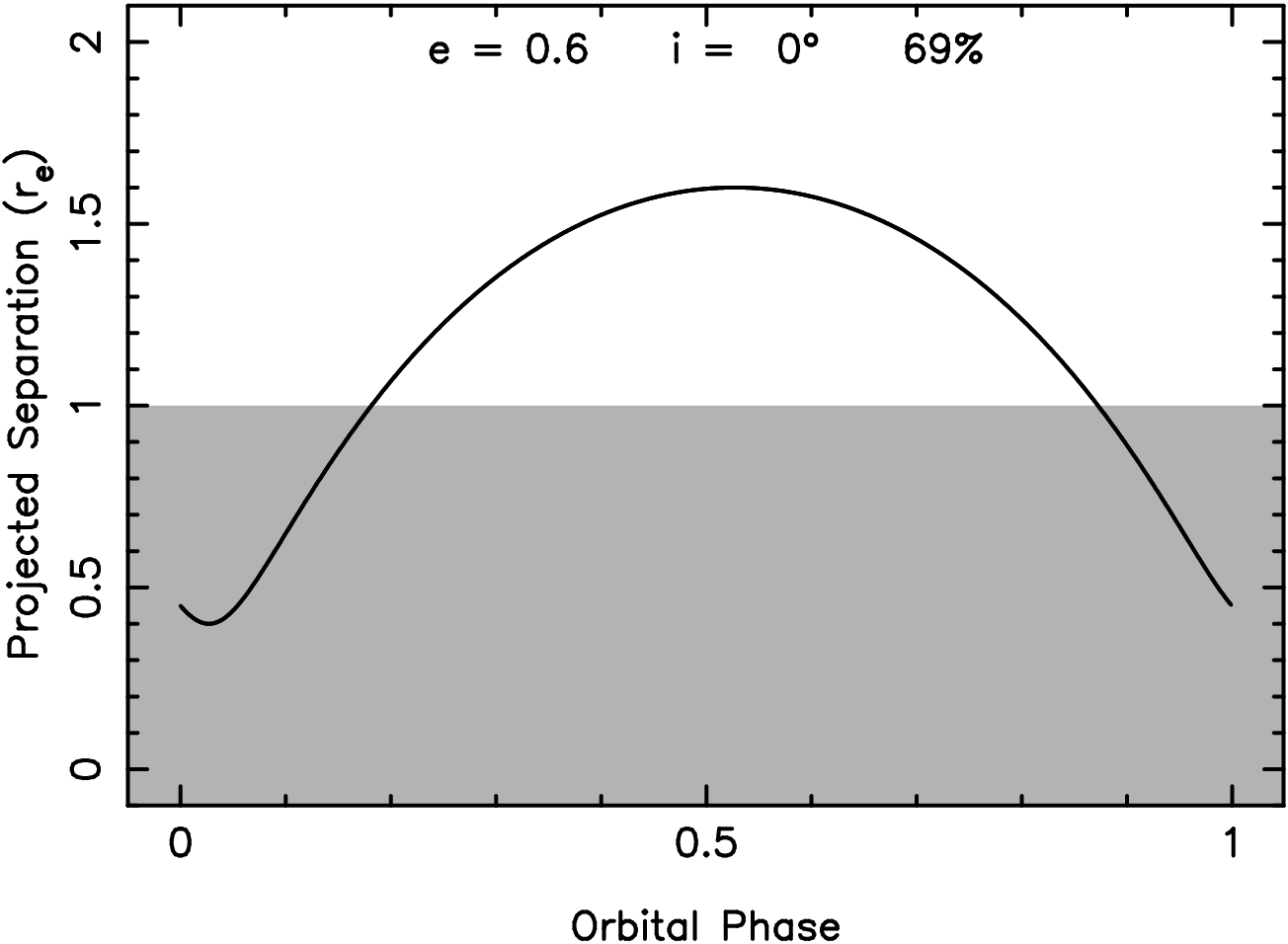} &
      \includegraphics[angle=270,width=5.2cm]{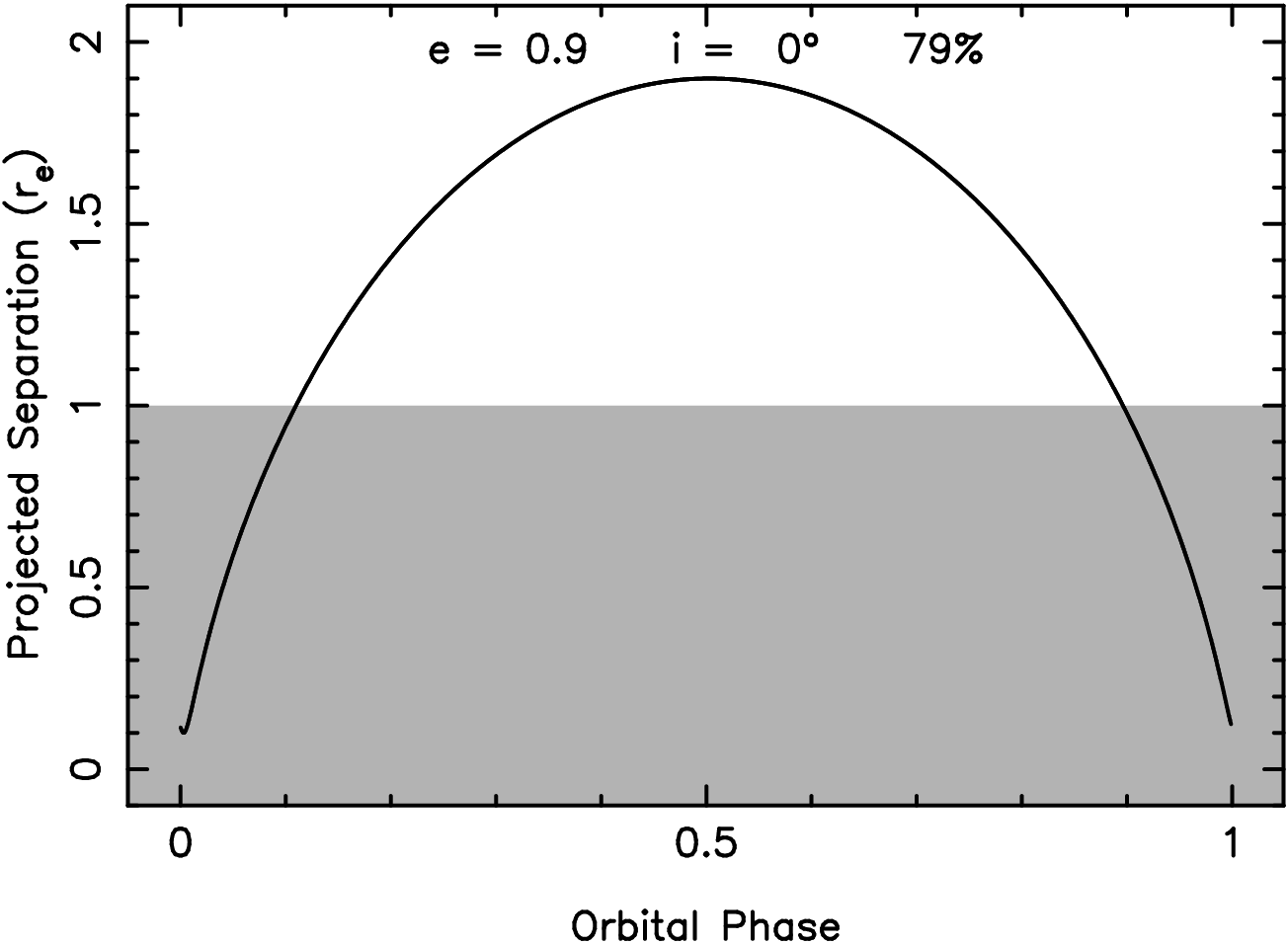} \\
      \includegraphics[angle=270,width=5.2cm]{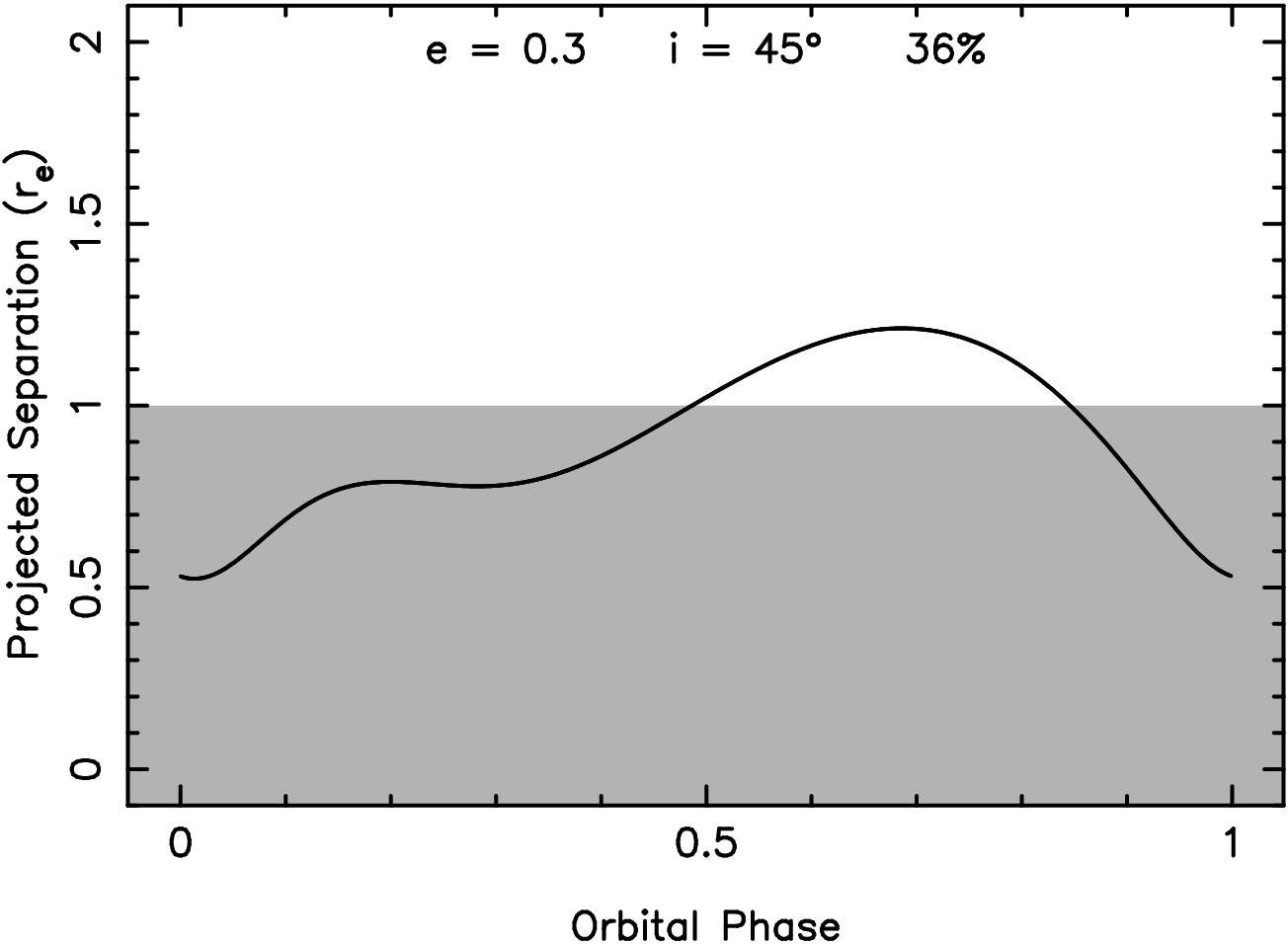} &
      \includegraphics[angle=270,width=5.2cm]{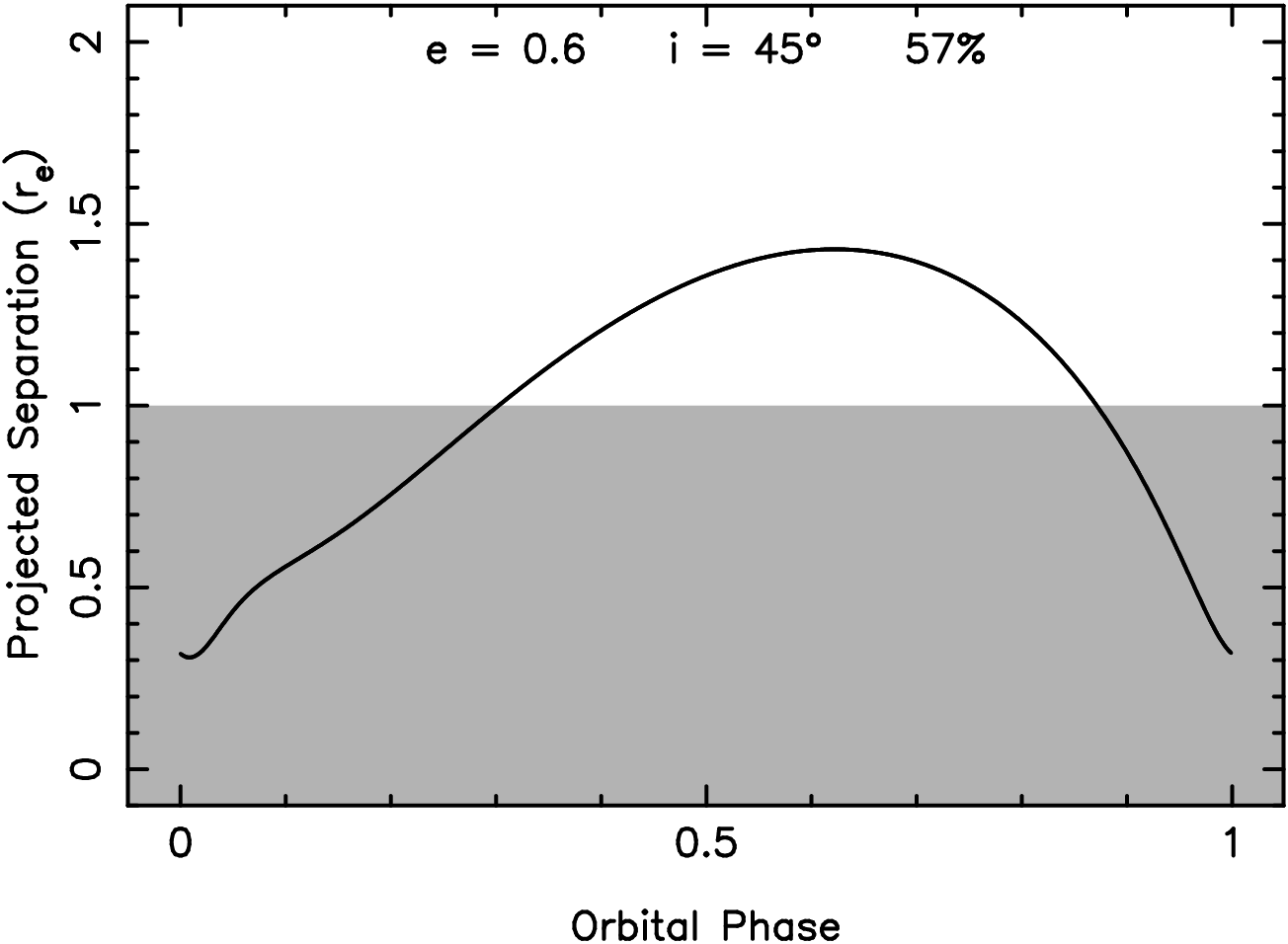} &
      \includegraphics[angle=270,width=5.2cm]{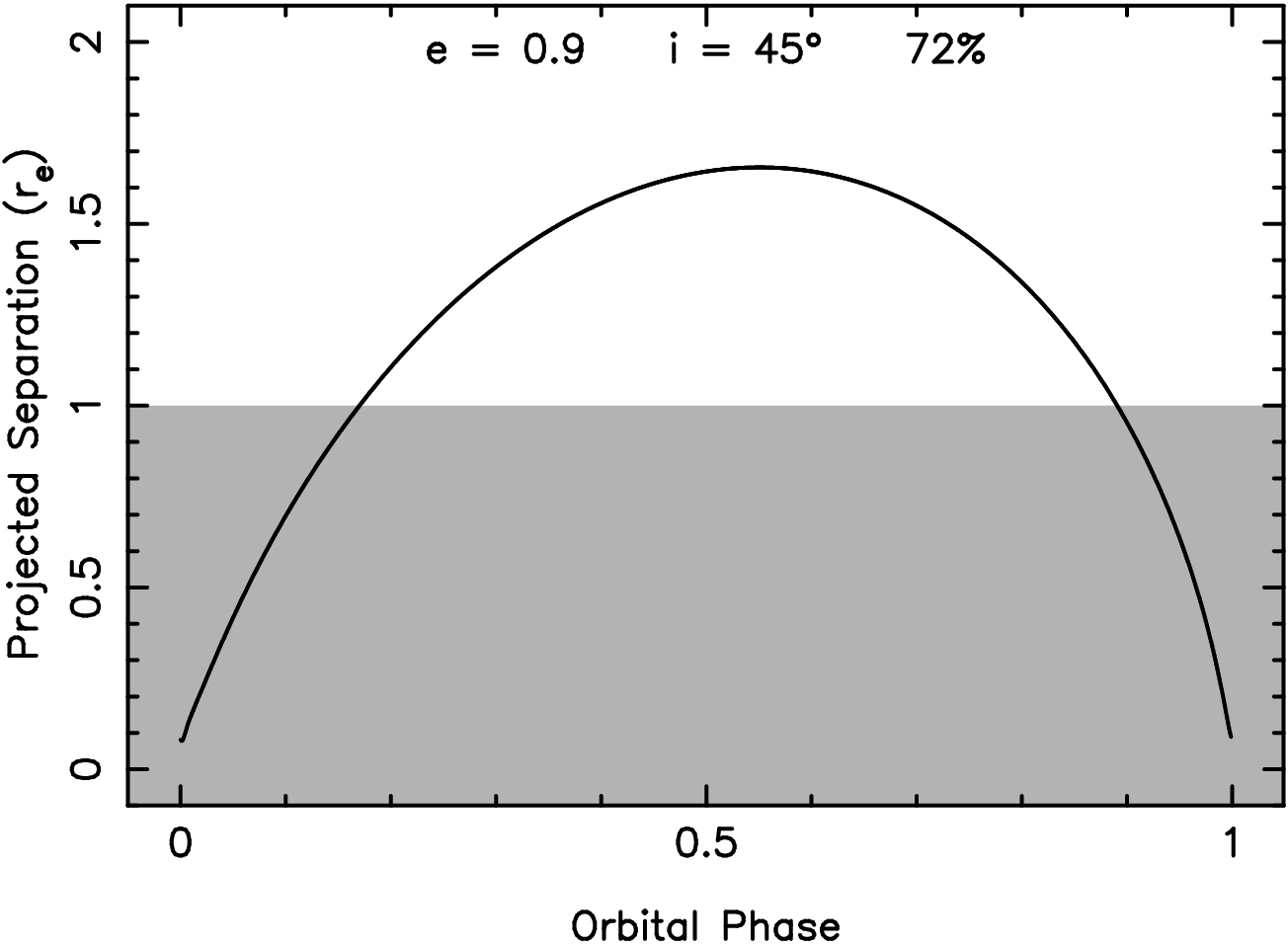} \\
      \includegraphics[angle=270,width=5.2cm]{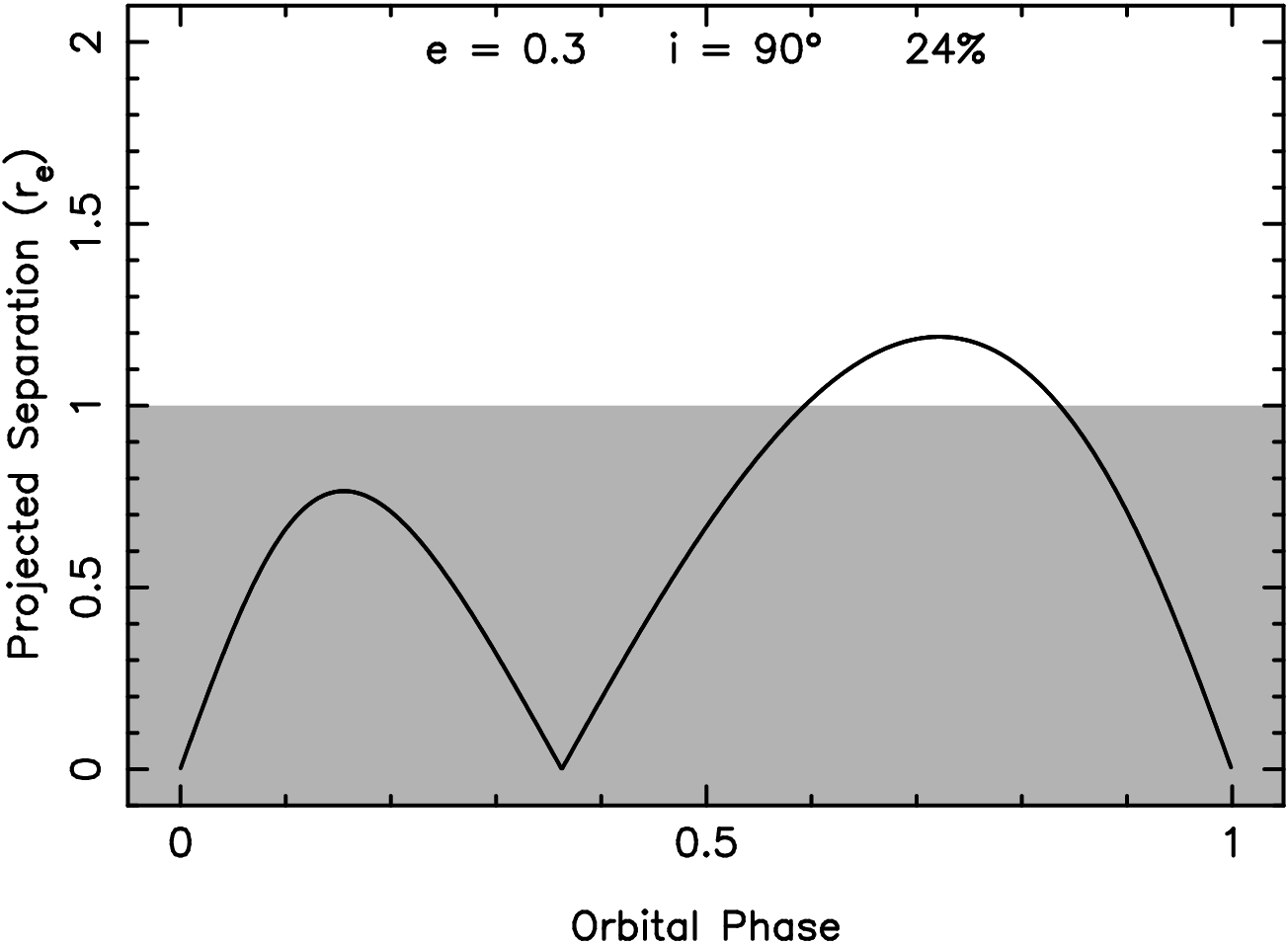} &
      \includegraphics[angle=270,width=5.2cm]{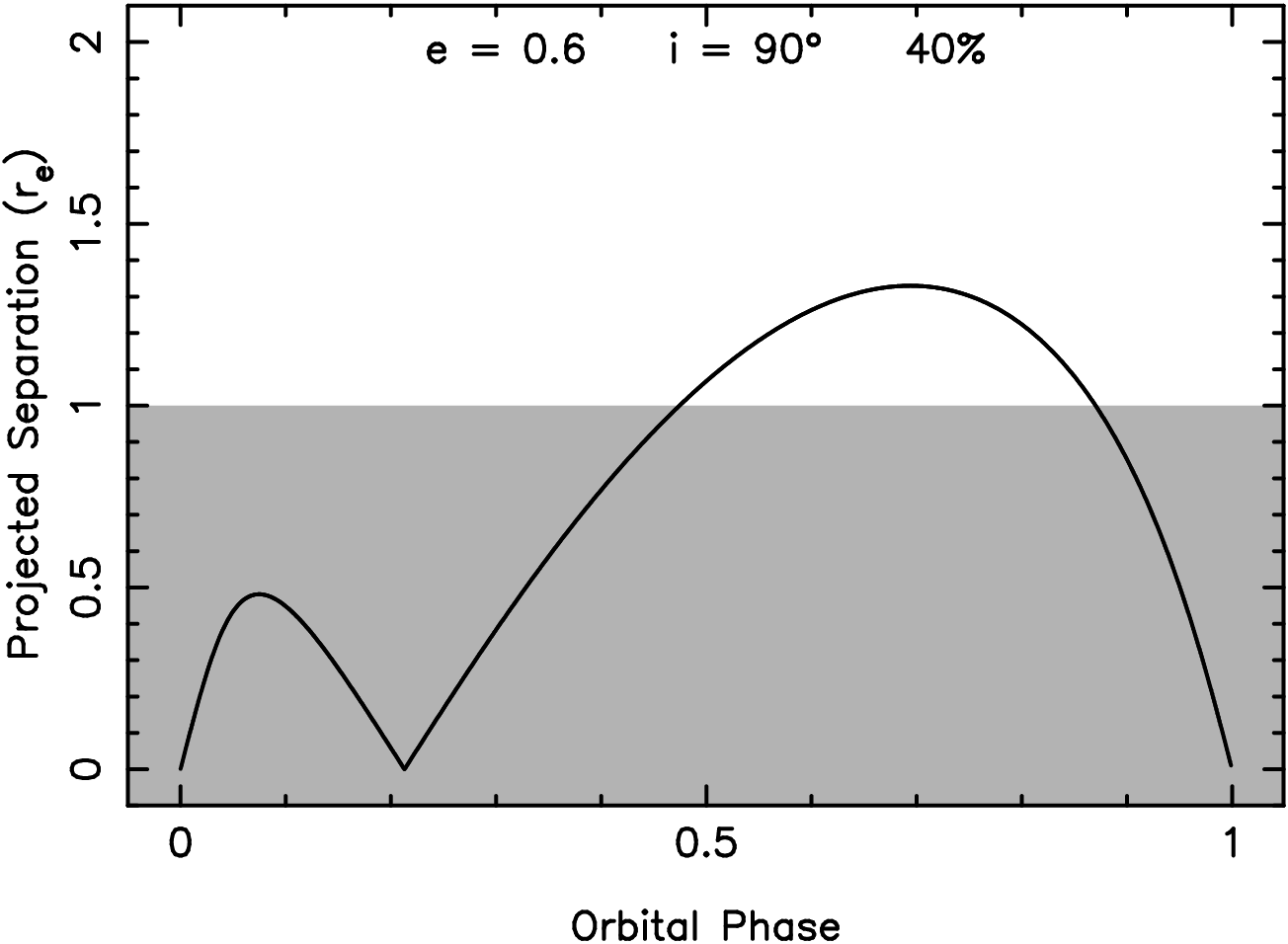} &
      \includegraphics[angle=270,width=5.2cm]{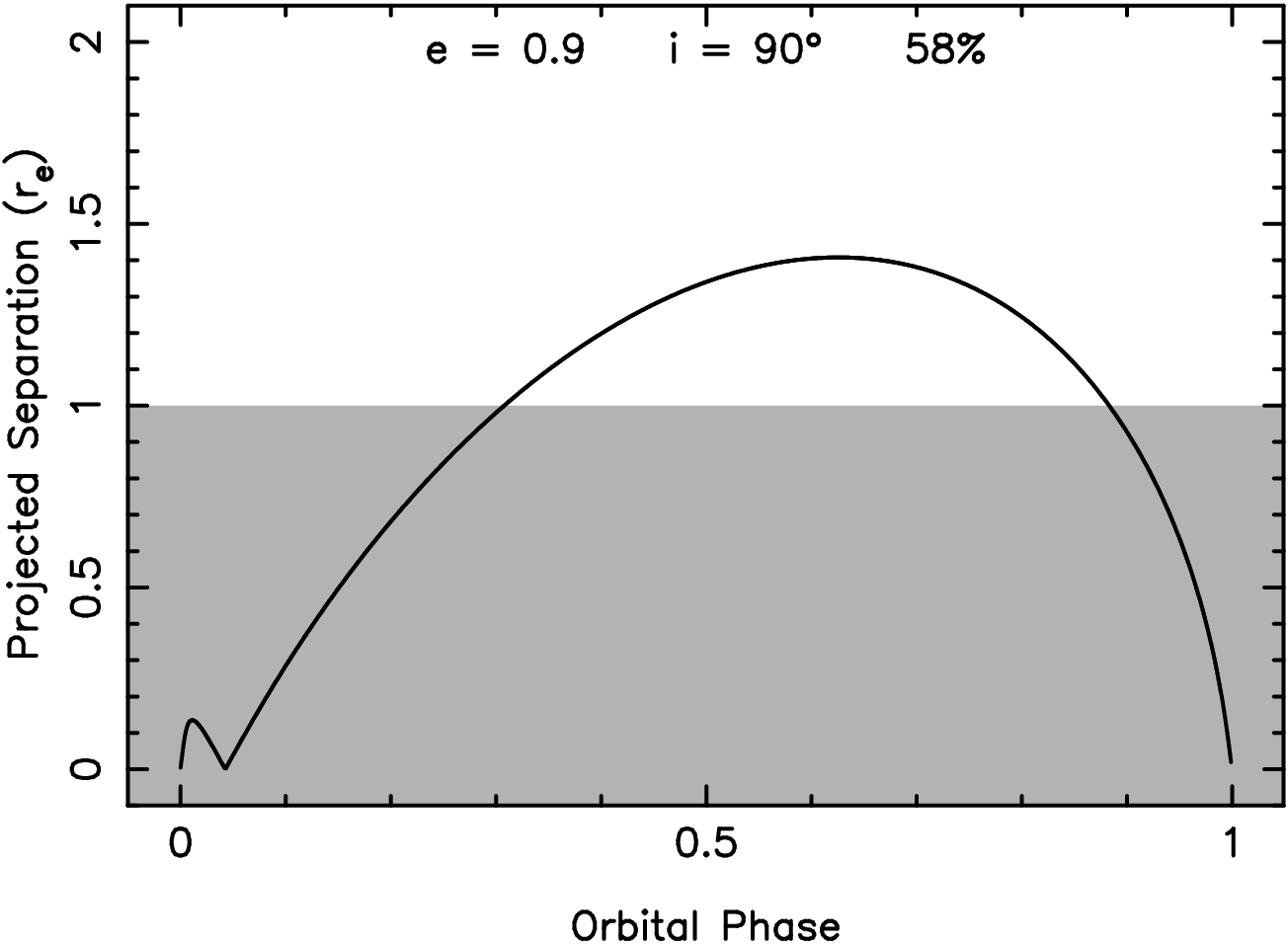}
    \end{tabular}
  \end{center}
  \caption{The projected star--planet separation as a function of
    orbital phase for eccentricities of $e = 0.3$, 0.6, and 0.9 and
    inclinations of $i = 0\degr$ (face-on), $45\degr$, and $90\degr$
    (edge-on). We use $\omega = 315\degr$ and $a = r_e$ in this
    example. The shaded region indicates the exclusion zone (as for
    Figure \ref{orbits} and the right-hand number in each plot is the
    percentage of the orbit that lies outside the exclusion zone.}
  \label{incarray}
\end{figure*}

We now utilize the components of the previous sections to calculate
the expected fraction of planets that will be detectable from images
experiments. To do this, we require the inclusion of the effects of
inclination since this can have a large effect on the projected
separation between the star and planet. Calculations from Equation
\ref{angsep} are combined with those of a Keplerian orbital solution
to compute the projected separation as a function of orbital phase,
where phase zero is the location in the orbit where $\omega + f =
270\degr$. We also set $\omega = 315\degr$ and $a = r_e$. The choice
$\omega$ is to demonstrate the asymmetry that can result when
considering eccentric orbits, a more thorough discussion with
application to photometric phase curves of which may be found in
\citet{kan11}. We show the results of these calculations in Figure
\ref{incarray} where we use three different values for eccentricity
($e = 0.3$, 0.6, and 0.9) and inclination ($i = 0\degr$ (face-on),
$45\degr$, and $90\degr$ (edge-on)). The shaded region indicates the
exclusion zone (as for Figure \ref{orbits}). The eccentricity,
inclination, and percentage of the orbit that lies outside the
exclusion zone, respectively, are given at the top of each subplot. In
general, the most favorable condition for detection is that for which
inclinations are close to face-on, although high-eccentricity orbits
suffer less in this regard when orbits become close to
edge-on. Depending on the asymmetry in the projected separation
(sensitively dependent on $\omega$), increased values of $a$ will
result in multiple opportunities for detection as the orbit moves in
and out of the exclusion zone.

\begin{figure}
  \includegraphics[angle=270,width=8.2cm]{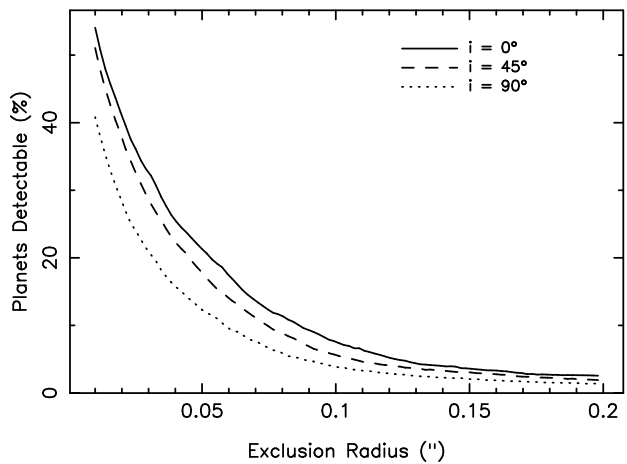}
  \caption{The results of a Monte-Carlo simulation which calculates
    the percentage of the known exoplanets (see Section \ref{eccdist})
    which would be detected using direct imaging as a function of the
    exclusion radius for that particular experiment. Since we cannot
    assume an inclination, we have performed this simulation for
    inclinations of $i = 0\degr$, $45\degr$, and $90\degr$.}
  \label{montecarlo}
\end{figure}

To quantify the effect of all these aspects on the detection
efficiency of imaging experiments, we performed a detail Monte-Carlo
simulation. This simulation calculates the percentage of the orbit
which lies outside the exclusion zone for each of the known exoplanets
described in Section \ref{eccdist}. We calculate this a function of
the exclusion zone expressed as an angular separation which uses the
known distance to the star. We also consider three inclinations for
all planets; $i = 0\degr$, $45\degr$, and $90\degr$. We then randomly
determine if the planet lies outside the exclusion zone, thus using
the percentage of time outside the exclusion zone as a probability
distribution. The results of this simulation are shown in Figure
\ref{montecarlo} where we have calculated the percentage of the
planets recovered in each of the simulations for each exclusion
radius. We consider the range of exclusion radii from 0.01\arcsec to
0.2\arcsec, a range for which many of the known exoplanets,
particularly those at short periods, will remain elusive. For
comparison, the angular separation of HR~8799~e from the host star is
0.37\arcsec \citep{mar10}, and the inner working angle of selected
future ground-based imaging instruments described by \citet{bei10}
range from 0.03\arcsec to 0.17\arcsec. The results of this simulation
support the conclusion from Figure \ref{incarray} that face-on
inclinations are more favorable for detection. More importantly
though, it demonstrates the dramatic dependence on the exclusion
radius for the known exoplanets. This dependence on the exclusion
radius is true even if all the orbits are circular. Note that assuming
an isotropic distribution of orbital inclinations by randomly
selecting the inclination for each planet in the Monte-Carlo
simulation results in a dependence that is roughly equivalent to the
$i = 45\degr$ case shown in Figure \ref{montecarlo}.

Finally, we performed a simulation which uses the observed
eccentricity distribution described in Section \ref{eccdist}. We first
used the eccentricities of the planets between 0.0--0.1~AU to
represent near circular orbits. This consists of 165 planets which we
use as the seed for the Monte-Carlo simulation whilst fixing all
semi-major axes to a given value. As for the previous simulation, we
choose random points in the orbital phase for each planet and
determine if the planet is detected (outside the exclusion radius) or
undetected (inside the exclusion radius). For semi-major axes of $a =
1.0 r_e$, 76\% of planets are detected, whereas for $a = 1.5 r_e$,
100\% of planets are detected. We then repeated this simulation by
using the eccentricities of the planets which lie between 1.0--2.0~AU;
a sample which is more representative of planets at longer orbital
periods where imaging experiments are more sensitive. For semi-major
axes of $a = 1.0 r_e$, 62\% of these planets are detected, and for $a
= 1.5 r_e$, 96\% of planets are detected. This is consistent with the
results shown in Figure \ref{percentage} and indicates that the vast
majority of eccentric planets with $a > 1.5 r_e$ should be detectable
at any given time. Note that this assumes that the eccentricity
distribution between 1.0--2.0~AU continues at longer orbital
periods. Figure \ref{exoplanets} suggests that the mean eccentricity
may decline beyond 2.0~AU. However, it is unclear if this is an
astrophysical reality or a symptom of RV survey incompleteness to
eccentric orbits at longer periods \citep{cum04,she08}.


\section{Discussion}
\label{discussion}

There are numerous imaging surveys which have considered the detection
limitations of their particular experimental designs
\citep{cha10,bon12,bra06,del12,jan12,laf07c,nie08,nie10,vig12}. These
often treat the eccentricity distribution as a basic gaussian function
based upon the RV exoplanets. Examples of the methods explored are the
commonly used probability density distribution approach described in
Appendix A of \citet{bra06} and the Monte Carlo simulation code (MESS)
  developed by \citet{bon12}. The difference with our approach here is
  that (a) we use the RV eccentricity distribution directly rather
  than infer a function from that distribution, and (b) we
  specifically determine the effect of eccentricity and the correlated
  parameters of periastron argument and inclination on detection
  efficiency.

The Monte-Carlo simulation we perform in Section \ref{deteff} uses the
eccentricity distribution of RV planets at long-periods as a proxy for
the expected distribution at even larger separations. However, as was
noted at the conclusion of the Section, it is unclear if this
distribution does indeed persist in this fashion. Additionally, the
samples of RV host stars and imaging survey targets tend to represent
quite different stellar demographics in terms of their distances and
ages. Eccentricity studies of binary stars by \citet{duq91} indicate
that the distribution peaks at $e \sim 0.35$ for $P > 1000$~days,
whereas the mean eccentricity of the known RV planets between
1.0--2.0~AU is 0.27 (see Section \ref{eccdist}). However, considering
the very different formation mechanisms between stars and planets,
such a comparison may be of dubious value. Constraints on the orbital
eccentricity of beta Pictoris b by \citet{cha12} and on the
eccentricities of the HR~8799 planets by \citet{cur12a} have shown
that these eccentricities are relatively low ($< 0.3$). Thus the
number of imaged planets are small enough and the eccentricities low
enough such that little may be discerned regarding the eccentricity
distribution from these planets alone. Increasing this sample by a
factor of 10 would start to produce meaningful insight into the
distribution of eccentricities beyond that which is currently being
probed by RV surveys. As imaging experiments are able to decrease the
exclusion radius and provide adequate observational cadence, the
ability to detect the orbital motion of detected planets (and thus
constrain the Keplerian properties of that orbit) will correspondingly
increase.


\section{Conclusions}

The exoplanet detection method of direct imaging is a powerful
technique which is rapidly developing in both technical and scientific
contexts. The major frontier push is to not only to detect smaller
masses, but to detect those at smaller separations from their
stars. Thus the overlap between the semi-major axis parameter-space of
the RV and imaging techniques is becoming larger at a rate such that
it's prudent to investigate the kinds of Keplerian orbital solutions
the imaging technique may expect to encounter with new
discoveries. Here we have shown the impact of orbital eccentricity on
exoplanet detectability through simulations, the observed distribution
of Keplerian orbital parameters, and the use of a simplistic model of
the exclusion zone for a particular experiment. In practice, the
exclusion zone is not well-defined and is a complex function of the
diffraction and speckle effects in the resulting images. The exclusion
zone used here though well represents an inner region where detections
are rendered virtually impossible by the flux of the star. These
results show that the eccentricity can have a profound effect of the
detection of eccentric exoplanets whose semi-major axis lie in the
range $0.5 < r_e < 1.5$. We have also quantified the effects of
inclination on these results and shown how face-on orbits can increase
the detection efficiency for eccentric orbits. As the techniques for
direct imaging improve and further epochs are acquired for the known
imaged systems, the precise effect of eccentricity on these
experiments will be become abundantly clearer and add enormously to
our knowledge of the eccentricity distribution as very long orbital
periods.


\section*{Acknowledgements}

The author would like to thank Thayne Currie for several enlightening
discussions on this topic, and also to Natalie Hinkel for providing
useful feedback on the manuscript. Thanks are also due to the
anonymous referee, whose comments improved the quality of the paper.
This research has made use of the Exoplanet Orbit Database and the
Exoplanet Data Explorer at exoplanets.org. This research has also made
use of the NASA Exoplanet Archive, which is operated by the California
Institute of Technology, under contract with the National Aeronautics
and Space Administration under the Exoplanet Exploration Program.


\end{document}